\documentstyle[twocolumn,pre,aps]{revtex}
\begin{document}
\draft

\title{An exact solution for the lattice gas model in one dimension}
\author{Juraj Vavro}
\address{Department of Material Science and Engineering, 
University of Pennsylvania, Philadelphia, 
Pennsylvania 19104
}
\maketitle

\begin{abstract}
A simple method to obtain a canonical partition function for one 
dimensional lattice gas model is presented. The simplification is 
based upon rewriting a sum over all possible configurations
to a sum over all possible numbers of clusters in the system.
\end{abstract}

\section{Canonical ensemble}

The behavior of canonical ensemble is fully described by its 
canonical partition function.  
In the case of the lattice gas model the partition function can be written in the form 

\begin{equation} 
   Z(L,N,\beta)=
   \sum_{\begin{array}{c} \sum_{i=1}^{L} e_i =N\end{array}}
         \!\exp(-E[\{e_i\}]\beta),
\end{equation}

where $L$ is the 
number of sites in the system, $N$ is the number of particles, 
$\beta=(k_BT)^{-1}$, $k_B$ being the Boltzman constant,
 $E[\{e_i\}]$ is 
potential energy of the system, $e_i=1$ if the $i$-th site is occupied
and $e_i=0$ otherwise. The main problem when calculating the canonical
partition function is accounting only for states
with total number of particles $N$,
that is to calculate degeneracies of the energy
levels.

Here I present a method 
for obtaining the partition function as well as clustering
probability for one-dimensional (1D) lattice gas model, 
assuming that the interaction between molecules is only
of nearest-neighbor type and that the binding energy 
$\varepsilon_s$ is constant 
for each site except for the leftmost site $i=1$ and the 
rightmost site $i=L$ where the binding energies are equal to
$\varepsilon_s+\varepsilon_L$ and 
$\varepsilon_s+\varepsilon_R$, respectively. 

First, it is useful to realize, that in the case of nearest neighbor 
interaction, the interaction energy depends only on the number of
clusters, and
not on their
 positions or their lengths. Then, the problem
of finding $Z(L,N,\beta)$ 
can be formulated as a problem of finding all possible
configurations of $k$ clusters. 

This is a unique property of one-dimensional systems and is based on the
fact that the boundary of a 1D cluster always consists of two points.
Two- and three-dimensional clusters do not posses this property; the 
boundary of two or three-dimensional clusters depend not only on the cluster
size but also on the configuration of particles within the cluster,
i.e., the cluster shape.

In the following it will be assumed that
$k\geq 1$, $L\geq N\geq 1$. 
It is suitable for further calculations to introduce function $F(k,n)$,
which counts the number of ways in which $k$ integer numbers
$a_i\geq 1 , i=1,\ldots,k$  can be chosen 
such that $\sum_{i=1}^{k}a_i=n$.
Thus, it is possible to write

\begin{equation}
 F(k,n)=\!\sum_{
              \begin{array}{c}
                       \sum_{i=1}^{k}a_i=n
              \end{array}
             } \!\!1 
=\sum_{m=1}^{n-k+1}F(k-1,n-m).
\end{equation}
This reccurent formula gives 

\begin{equation}
F(k,n)=
       \left\{
       \begin{array}{cl}
              \left(\begin{array}{c} n-1 \\ k-1
                                 \end{array}
                           \right) & \mbox{if $k\geq 1 \wedge n\geq k$}
\\
0 & \mbox{otherwise.}
\end{array} \right.
\end{equation}

Thus, the number of ways to 
arrange $N$ particles into $k$ clusters is given by $F(k,N)$
and the number of ways to arrange $(L-N)$ empty sites into 
$k'$ clusters is given by $F(k',L-N)$. So that the number of configurations
of the system with $N$ occupied sites arranged into $k$ clusters
and $(L-N)$ epty sites arranged into $k'$ clusters can be expressed
as $F(k,N)F(k',L-N)$.

Now it is suitable to express the canonical partition function
$Z(L,N,\beta)$ as a sum over number of clusters $k=1,\ldots,k_{max}$,
$k_{max}$ being equal to $\min(L-N+1,N)$

\begin{equation}
Z(L,N,\beta)=\sum_{k=1}^{k_{max}}
\!Z_{cl}(L,N,k,\beta)\exp((N-k)
\varepsilon\beta),
\label{eq:Z}
\end{equation}
where $Z_{cl}(L,N,k,\beta)$ is the partition function for
a system of length $L$ with $N$ particles in $k$ clusters and $k_{max}
$ gives the maximum number of clusters the system can contain.

\subsection{System with interaction between particles and boundaries}

To calculate the canonical partition 
function of the system $Z(L,N,\beta)$ 
in 
the case of interaction 
of particles with the 
boundaries of the sytem, 
it is convenient to write the function
$Z_{cl}(L,N,k,\beta)$ as a sum of three
parts

\begin{equation}
Z_{cl}(L,N,k,\beta)=\sum_{i=0}^{2}W_{i}(L,N,k)\exp({\mathcal{E}}_{i}
\beta),
\label{eq:Zcl}
\end{equation}
where 
\begin{eqnarray}
{\mathcal{E}}_2&=&\varepsilon_R+\varepsilon_L, \nonumber \\
{\mathcal{E}}_1&=&{\beta}^{-1}\log(\exp(\varepsilon_R\beta)+
                 \exp(\varepsilon_L\beta)), \nonumber \\
{\mathcal{E}}_0&=&0,
\end{eqnarray}
and
$\varepsilon_{R}$($\varepsilon_L$)
 is nearest neighbor interaction 
between right (left) end of the system and particle,
and $W_i(L,N,k), i=0,1,2$ are the 
numbers of realizations of system with the 
length $L$ containing 
$N$ particles arranged into $k$ clusters and with
$i$ clusters attached to the ends of the system.

In the case when sites $1$ and $L$ are occupied (two clusters
attached to the ends of the system,$i=2$), 
 we have $k'=(k-1)$ clusters of $(L-N)$ empty sites 
and $k$ clusters of $N$ particles
so that we can write

\begin{equation}
W_{2}(L,N,k)=\delta_{L,N}+F(k,N)F(k-1,L-N),
\label{eq:W2}
\end{equation}
where
\begin{equation}
\delta_{n,m}=\left\{\begin{array}{ll} 1 & \mbox{if $n=m$} \\
                                  0 & \mbox{otherwise.}
               \end{array}\right.
\end{equation}

When only one of the sites $1,L$ is occupied ($i=1$), then 
 $(L-N)$ empty sites  are arranged into $k'=k$ clusters
and again we have $k$ clusters of $N$ particles and we can
write
\begin{equation}
W_1(L,N,k)=
F(k,N)F(k,L-N).
\label{eq:W1}
\end{equation}

When neither one of the sites $1,L$ ($i=0$) is occupied 
we are dealing with
$k'=(k+1)$ clusters of $(L-N)$ empty sites and
$k$ clusters of N particles, so that it is possible to write
\begin{equation}
W_0(L,N,k)=
F(k,N)F(k+1,L-N).
\label{eq:W0}
\end{equation}

Thus, using the equation~(\ref{eq:Z}) together with the equations~(\ref
{eq:Zcl},\ref{eq:W2},\ref{eq:W1}) and (\ref{eq:W0}) we can calculate
the partition function of the system together with
all thermodynamics quantities like
mean energy of the system 
$E(L,N,T)$, thermal capacity $C(L,N,T)$,
pressure $p(L,N,T)$ and  probability 
$P_{cl}(L,N,\beta,k)$  
to find $k$ clusters in the system, which can be written
in the form 
\begin{equation}
P_{cl}(L,N,\beta,k)=\frac{Z_{cl}(L,N,k,\beta)}{Z(L,N,\beta)}
\exp\left((N-k)\varepsilon\beta\right).
\end{equation}

To calculate the probability $P_{p}(N,L,\beta,x)$
to find a cluster of length $x$ 
it is needed to start with 
 $W_p(N,L,k,k',x)$, which is the number of
clusters of the length $x$ when taking into account all possible
configurations of system of the length $L$ with $N$ 
particles arranged into $k$ clusters and with $(L-N)$ empty sites
arranged into $k'$ clusters. 
It can be  showed that for $k\geq 2$
\begin{eqnarray}
&&W_p(N,L,k,k',x)=kF(k-1,N-x)F(k',L-N),
\end{eqnarray}
and for $k=1$ we get
\begin{equation}
W_p(N,L,1,k',x)=\delta_{N,x}\bigl(F(k',L-N)+\delta_{L,N}\bigr).
\end{equation}

Using the same arguments and method that were used to calculate
$Z(L,N,\beta)$, it is possible to express $P_{p}(L,N,\beta,x)$ as
\begin{eqnarray}
\lefteqn{P_p(L,N,\beta,x)=
\frac{1}{Z_p(L,N,\beta)}} \nonumber \\
&&\times\Biggl[Z_{cl}(L,N,1,\beta)\delta_{N,x}\exp(-\varepsilon\beta)
\nonumber \\ &\ &+
\sum_{k=2}^{k_{max}}
kZ_{cl}(L,N,k,\beta)\frac{F(k-1,N-x)}{F(k,N)}
\exp(-k\varepsilon\beta)
\Biggr]\nonumber \\ &\ &\ 
\end{eqnarray}

It is important to point out that $Z_{p}(L,N,\beta)$ and
$Z(L,N,\beta)$ are different functions: $Z_{p}(L,N,\beta)
$ is  normalization
factor needed to satisfy condition $\sum_{x=1}^{N}P_p(L,N,\beta,x)=1$
, while $Z(L,N,\beta)$ is the partition function for the system of 
$L$ sites containing $N$ particles.

\subsection{Periodic boundary conditions}

The calculation of the canonical partition function 
$Z_{pbc}(L,N,\beta)$ in the case of the periodic
boundary conditions 
is now very straightforward, it is enough
to realize that this problem is equivalent to the case of ends 
interacting with particles with interaction energies 
$\varepsilon_R=\varepsilon_L=\varepsilon/2$
when both first and last site are occupied and 
$\varepsilon_R=\varepsilon_L=0$
if one of these sites is empty.
It immediately yields result

\begin{equation}
Z_{pbc}(L,N,\beta)=\sum_{k=1}^
{k_{max}}Z_{cl}^{(pbc)}(L,N,k,\beta)\exp((N-k)
\varepsilon\beta),
\end{equation}
where
\begin{equation}
Z_{cl}^{(pbc)}(L,N,k,\beta)=\delta_{L,N}
\exp(\varepsilon\beta
)+
\frac{L}{k}F(k,L-N)F(k,N).
\label{eq:pbc}
\end{equation}
Due to the periodic boundary conditions the maximum number of clusters
$k_{max}$ the system can contain is now equal to $\min(L-N,N)$.

The expression~\ref{eq:pbc} can be easily understood. The first part
$\delta_{L,N}\exp(\varepsilon\beta)$ is trivial and it represents a
special case when $L=N$, i.\ e.\ every site occupied. If $L>N$, then
due to the periodic boundary conditions the number of clusters of
particles is equal to the number of clusters of empty sites. Thus,
$F(k,N)$ counts the number of ways to arrange $N$ particles into
$k$ clusters and $F(k,L-N)$ does the same with $L-N$ empty sites.
Factor $L$ arises from the fact, that there are $L$ ways how to
position each arrangement of clusters, and factor $1/k$ is there
because in the system with the periodic boundary conditions
we do not know which cluster is first 
- and there are exactly $k$ ways to choose the first.

Calculating the probability \nolinebreak
$P_{cl}^{(pbc)}(L,N,\beta,k)$ to find $k$ clusters in the system 
is now easy
\begin{equation}
P_{cl}^{(pbc)}(L,N,\beta,k)=\frac{
Z_{cl}^{(pbc)}(L,N,k,\beta)}
{
Z_{pbc}(L,N,\beta)
}
\exp((N-k)\varepsilon\beta),
\end{equation}
and
the probability $P_{p}^{(pbc)}(L,N,\beta,x)$ 
to find cluster of length $x$ can be 
written in the form
\begin{equation}
P_{p}^{(pbc)}(L,N,\beta,x)=
\frac{
\sum_{j=0}^{2}W_{j}^{(pbc)}(L,N,\beta,x)
\exp(j\varepsilon\beta)
}
{Z_{p}^{(pbc)}(L,N,\beta)},
\end{equation}
where $Z_{p}^{(pbc)}(L,N,\beta)$ is normalization factor 
such that $\sum_{x=1}^{N}P_{p}^{(pbc)}(L,N,\beta,x)=1$
and $W_j,j=0,1,2$ can be expressed as
\begin{equation}
W_{0}^{(pbc)}
(L,N,\beta,x)=L\Theta(L-N-1)\delta_{x,N} + \exp(\varepsilon)
\delta_{x,L},
\end{equation}
\begin{eqnarray}
\lefteqn{
W_{1}^{(pbc)}(L,N,\beta,x)=\sum_{k=2}^{k_{max}}
kF(k-1,N-x)} \nonumber \\ 
&&\times
\bigl[F(k+1,L-N)
+F(k,L-N)\bigr]
\exp(-k\varepsilon\beta),
\end{eqnarray}
and
\begin{eqnarray}
\lefteqn{
W_{2}^{(pbc)}(L,N,\beta,x)=\sum_{k=3}^{k_{max}}
F(k-1,L-N)
}
\nonumber \\
&\times&\bigl[ (k-2)F(k-1,N-x)+(x-1)F(k-2,N-x)\bigr]
\nonumber \\
&\times&
\exp
\bigl(-k\varepsilon\beta\bigr).
\end{eqnarray}

 \section{Grandcanonical ensemble}
 
It is now possible to calculate properties
of the open system (grand-canonical ensemble) 
since the grand-canonical partition
function can be writen as a sum of canonical partition functions over
number of particles

\[
Q(L,\beta,\mu)=1+\sum_{N=1}^{L}Z(L,N,\beta)\exp(N(\varepsilon_{s}+
\mu)\beta)
\]

where $\varepsilon_{s}$ is an interaction energy between particle
and site (in our 1D system) different from $1,L$. However, in the case
of the periodic boundary conditions or in the case of the noninteracting
boundaries it is simpler to use the transfer matrix method~\cite{matrix}
.

\section{Conclusions}

A simple method for calculation of the  properties of
open and closed 1D systems with the periodic boundary 
conditions and with the boundaries interacting with particles
was presented.
Numerical values can be easily obtained,
since the summations over all possible configurations of
$N$ particles  on $L$ sites were rewritten to the
summations over $k=1,\ldots,\min(L-N+1,N)$, where
$\min(L-N+1,N)$ is the maximum number of clusters in the system,
this number being considerably smaller than the number of
all
possible configurations of the system. 

The method employs the fact that the boundary of each cluster
consists of two points. This is a unique property of one dimension, so that
it is not possible to generalize this method to higher
dimensions.

The results can be used in the study of equilibrium
properties of quasi-1D clusters. An ideal system on which
it is possible to study low-dimensional physics is
provided by single-walled carbon tubes (SWNT's).
These can be filled with various molecules and atoms, and if the
radius of a SWNT's is sufficiently small, the motion of trapped particles
is confined along the tube axis. What makes SWNT's  special
is the possibility to open and close the ends of nanotubes
\cite{Kuznetsova}, providing the opportunity to study both open and closed
systems.  
Recently, the presented method was succesfully used by Hodak and Girifalco \cite{hodo}
for a theoretical study of clustering 
$\mathrm{C}_{60}$ molecules encapsulated
in carbon nanotubes. This type of system, also
called peapod, 
was experimentaly observed by Smith
et al.\ \cite{brian}.

Other interesting systems, which can be used to probe the
physics of low-dimensional structures and for which the
obtained results may be applicable are provided by the interstitial
channels in ropes of SWNT's \cite{Stan,Teizer} and by the grooves
on the external surface of the ropes \cite{Williams}, both of which
can provide high-energy adsorption sites.

The presented method can also be useful in analysis of the stability
of one-dimensional metals, which can be obtained for example
by alkali-metal absorption on semiconductor surfaces \cite{Modesti},
and in the study of the behaviour of classical gases confined
in nanopore materials \cite{Ebbesen} and zeolite materials
\cite{Stan}.

\acknowledgments

The author is thankful to Professor John E.\ Fischer for
his continuous support and help. The author also
gratefully ackonowledges support through U.\ S.\
Department of Energy, DEFG02-98ER45701.
\newline

\end{document}